\documentclass[aps,prl,floatfix,twocolumn,showpacs,10pt]{revtex4-1}

\usepackage{graphicx}
\usepackage{dcolumn}
\usepackage{bm}
\usepackage{amssymb}
\usepackage{rotating} 

\begin{document}


\title{Structure of Fermionic Density Matrices: Complete $N$-representability Conditions}

\author{David A. Mazziotti}

\email{damazz@uchicago.edu}
\affiliation{Department of Chemistry and The James Franck Institute, The University of Chicago, Chicago, IL 60637}%

\date{Submitted November 8, 2011; Published {\em Phys. Rev. Lett} {\bf 108}, 263002 (2012)}


\begin{abstract}

We present a constructive solution to the $N$-representability
problem---a full characterization of the conditions for constraining
the two-electron reduced density matrix (2-RDM) to represent an
$N$-electron density matrix. Previously known conditions, while
rigorous, were incomplete. Here we derive a hierarchy of constraints
built upon (i) the bipolar theorem and (ii) tensor decompositions of
model Hamiltonians. Existing conditions $D$, $Q$, $G$, $T1$, and
$T2$, known classical conditions, and new conditions appear
naturally. Subsets of the conditions are amenable to polynomial-time
computations of strongly correlated systems.

\end{abstract}


\maketitle

The wavefunction of a many-electron quantum system contains
significantly more information than necessary for the calculation of
energies and properties.  In 1955 Mayer proposed in {\em Physical
Review} computing the ground-state energy variationally as a
functional of the two-electron reduced density matrix (2-RDM) which,
unlike the wavefunction, scales polynomially with the number $N$ of
electrons~\cite{RDM,CY00,M55}.  However, the 2-electron density
matrix must be constrained to represent a many-electron (or
$N$-electron) density matrix (or wavefunction); otherwise, the
minimized energy is unphysically below the ground-state energy for
$N>2$. Coleman called these constraints {\em $N$-representability
conditions}~\cite{C63}, and the search for them became known as the
$N$-representability problem~\cite{GP64,H78,E78,E79,P78,R07}. In
1995 the National Research Council ranked the $N$-representability
problem as one of the top unsolved theoretical problems in chemical
physics~\cite{NRC}.  While progress was limited for many years,
recent advances in theory and optimization~\cite{EJ00,N01,M04,
P04,C06,E07,A09,S10,M11} have enabled the application of the
variational 2-RDM method to studying strong correlation in quantum
phase transitions~\cite{GM06}, quantum dots~\cite{RM09},
polyaromatic hydrocarbons~\cite{GM08}, firefly
bioluminescence~\cite{GM10}, and metal-to-insulator
transitions~\cite{SGM10}.

Despite the recent computational results with 2-RDM methods, a
complete set of $N$-representability conditions on the 2-RDM---not
dependent upon higher-order RDMs---has remained unknown.  While
formal solutions of the $N$-representability problem were developed
in the 1960s~\cite{GP64,K67}, practically they required the
$N$-electron density matrix~\cite{RDM,CY00}.  In this Letter we
present a constructive solution of the $N$-representability problem
that generates a complete set of $N$-representability conditions on
the 2-RDM.  The approach is applicable to generating the
$N$-representability conditions on the $p$-RDM for any $p \le N$.
The conditions arise naturally as a hierarchy of constraints on the
2-RDM, which we label the $(2,q)$-positivity conditions, where the
$(2,2)$- and $(2,3)$-positivity conditions include the already known
$D$, $Q$, $G$, $T1$, and $T2$ conditions~\cite{C63,GP64,E78,P04}.
The second number in $(2,q)$ corresponds to the higher $q$-RDM which
serves as the starting point for the derivation of the condition.

A key advance in extending the $(2,q)$-positivity conditions for
$q>3$ is the use of tensor decompositions in the model Hamiltonians
that expose the boundary of the $N$-representable 2-RDM set.  The
decompositions allow the terms in the model Hamiltonians to have no
more than two-body interactions through the cancelation of all
higher 3-to-$q$-body terms.  A second important element is the
recognition that when $q=r$ where $r$ is the rank of the
one-electron basis set the positivity conditions are complete. The
hierarchy of conditions can be thought of as a collection of model
Hamiltonians~\cite{P78}. For example, the `basic' (2,2)-positivity
conditions are both necessary and sufficient constraints for
computing the ground-state energies of pairing model
Hamiltonians~\cite{CY00,M04}, often employed in describing
long-range order and superconductivity.

Consider a quantum system composed of $N$ fermions.  A matrix is a
fermionic {\em density matrix} if and only if it is: ({\em i})
Hermitian, ({\em ii}) normalized (fixed trace), ({\em iii})
antisymmetric in the exchange of particles, and ({\em iv}) positive
semidefinite. A matrix is {\em positive semidefinite} if and only if
its eigenvalues are nonnegative.  The $p$-particle reduced density
matrix ($p$-RDM) can be obtained from the $N$-particle density
matrix by integrating over all but the first $p$ particles
\begin{equation}
\label{eq:Dp} {}^{p} D = {N \choose p} \int{ {}^{N} D \, d(p+1)
\dots dN } .
\end{equation}
The set of ${}^{N} D$ is a convex set which we denote as $P^{N}$
while the set ${}^{p} D$ is a convex set which we denote as
$P^{p}_{N}$, the set of $N$-representable $p$-particle density
matrices.  A set is {\em convex} if and only if the convex
combination of any two members of the set is also contained in the
set
\begin{equation}
w \, {}^{N} D_{1} + (1-w) \, {}^{N} D_{2} \in P^{N},
\end{equation}
where $0 \le w \le 1$.  The integration in Eq.~(\ref{eq:Dp}) defines
a linear mapping from $P^{N}$ to $P^{p}_{N}$, which preserves its
convexity.

The energy of a quantum system in a stationary state can be computed
from the Hamiltonian traced against the state's density matrix.  For
a system of $N$ fermions we have
\begin{equation}
\label{eq:EN} E = {\rm Tr}({\hat H} \, {}^{N} D) .
\end{equation}
If the Hamiltonian is a $p$-body operator, meaning that it has at
most $p$-particle interactions, then the energy can be written as a
functional of only the $p$-RDM
\begin{equation}
\label{eq:Ep} E = {\rm Tr}({\hat H} \, {}^{p} D) .
\end{equation}
For a system of $N$ electrons the Hamiltonian generally has at most
pairwise interactions, and hence, the energy can be expressed as a
linear functional of the 2-RDM.  Except when $N=2$, however,
minimizing the energy as a functional of a two-electron density
matrix ${}^{2} D \in P^{2}$ yields an energy that is much too low.
To obtain the correct ground-state energy, we must constrain the
two-electron density matrix to be $N$-representable, that is $^{2} D
\in P^{2}_{N}$.

Based on the equivalence of the energy expectation values in
Eqs.~(\ref{eq:EN}) and~(\ref{eq:Ep}), we can use the set $P^{p}_{N}$
of $N$-representable $p$-particle density matrices to define a set
${P^{p}_{N}}^{*}$ of $p$-particle (Hamiltonian) operators $^{p}
{\hat O}$ that are positive semidefinite in their trace with any
$N$-particle density matrix
\begin{equation}
\label{eq:Oset} {P^{p}_{N}}^{*} = \{ ^{p} {\hat O} | {\rm Tr}(^{p}
{\hat O} \, ^{p} D) \ge 0~{\rm for~all}~^{p} D \in {P^{p}_{N}} \}.
\end{equation}
The set ${P^{p}_{N}}^{*}$ is said to be the {\em polar} (or dual) of
the set $P^{p}_{N}$. Importantly, by the {\em bipolar
theorem}~\cite{K67,R71}, the set ${P^{p}_{N}}^{*}$ also fully
defines its polar set $P^{p}_{N}$ as follows
\begin{equation}
\label{eq:Dset} P^{p}_{N} = \{ ^{p} D | {\rm Tr}(^{p} {\hat O} \,
^{p} D) \ge 0~{\rm for~all}~^{p} {\hat O} \in {{P^{p}_{N}}^{*}} \}.
\end{equation}
By Eq.~(\ref{eq:Dset}) we have a complete characterization of the
$N$-representable $p$-RDMs from a knowledge of all operators $^{p}
{\hat O} \in {P^{p}_{N}}^{*}$~\cite{K67}.  This analysis shows
formally that there exists a solution to the $N$-representability
problem~\cite{GP64,K67}, but it does not provide a mechanism for
characterizing the set ${P^{p}_{N}}^{*}$.


\begin{figure}[htp!]

\includegraphics[scale=1.0]{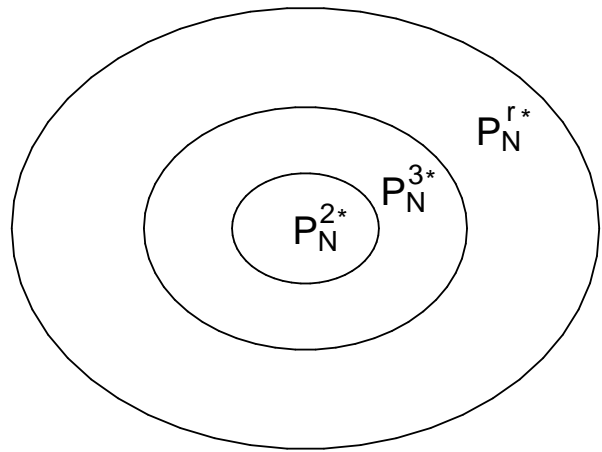}

\caption{The convex set ${P^{2}_{N}}^{*}$ of 2-body operators that
are positive semidefinite in their trace with any $N$-particle
density matrix is contained within the convex set ${P^{3}_{N}}^{*}$
of analogous 3-body operators, which in turn is contained within the
set ${P^{r}_{N}}^{*}$.  Hence, the extreme points of
${P^{2}_{N}}^{*}$ can be characterized completely by the convex
combination of the extreme points of ${P^{r}_{N}}^{*}$, which are
given by Eq.~(\ref{eq:rpos}).}

\label{f:n2}

\end{figure}

To characterize ${P^{p}_{N}}^{*}$, we assume that the $N$-fermion
quantum system has $r$ orbitals and hence, $r-N$ holes.  A convex
set can be defined by the enumeration of its {\em extreme elements},
that is the elements (or members) that cannot be expressed by a
convex combination of other elements~\cite{CY00,R71}.  The
definition of ${P^{p}_{N}}^{*}$ in Eq.~(\ref{eq:Oset}) for $p \le N$
can be extended in second quantization to include $p > N$
\begin{equation}
\label{eq:Oset2} {P^{p}_{N}}^{*} = \{ ^{p} {\hat O} | {\rm Tr}(^{p}
{\hat O} \, ^{N} D) \ge 0~{\rm for~all}~^{N} D \}
\end{equation}
with the $^{p} {\hat O}$ being polynomials in creation and
annihilation operators of degree $2p$.  Because in second
quantization the value of $N$ is defined in the density matrices
$^{N} D$ rather than in the operators $^{p} {\hat O}$~\cite{S89},
the set ${P^{p}_{N}}^{*}$ provides complete $N$-representability
conditions on the $p$-RDM for any $N$ between 2 and $r$.  The
extreme operators in the set ${P^{r}_{N}}^{*}$ can be written as
Hermitian squares of operators~\cite{H02}
\begin{equation}
\label{eq:rpos} {^{r} {\hat O}_{i}} = {^{r} {\hat C}_{i}} \, {^{r}
{\hat C}_{i}^{\dagger}},
\end{equation}
where the ${}^{r} {\hat C}_{i}$ are polynomials in the creation and
annihilation operators of degree less than or equal to $r$ (i.e.,
Eqs.~(\ref{eq:T21}) and (\ref{eq:T22})). Because any operator
${}^{p} {\hat C}$ with $p>r$ reduces to a polynomial of degree $r$
in its operation on any ${}^{N} D$, the sets ${P^{p}_{N}}^{*}$ with
$p>r$ do not contain additional information about the positivity of
${}^{N} D$.  To establish this reduction, we rearrange terms in
${}^{p} {\hat C}$ of degree greater than $r$ into a normal order
with either more than $N$ annihilation operators to the right of the
creation operators or more than $r-N$ creation operators to the
right of the annihilation operators; in either situation, the terms
of degree greater than $r$ vanish in their operation upon any $^{N}
D$.

The operators ${}^{p} {\hat O}$ that constrain the $p$-RDM to be
$N$-representable in Eq.~(\ref{eq:Dset}) are also necessary to
constrain the $q$-RDM to be $N$-representable where $q>p$; formally,
each ${}^{p} {\hat O} \in {P^{p}_{N}}^{*}$ can be lifted by
inserting the number operator to the $(q-p)$ power to form a ${}^{q}
{\hat O} \in {P^{q}_{N}}^{*}$~\cite{M04}. Therefore, as illustrated
in Fig.~1, we have the following set relations
\begin{equation}
{P^{2}_{N}}^{*} \subseteq {P^{3}_{N}}^{*} \subseteq {P^{p}_{N}}^{*}
... \subseteq {P^{r}_{N}}^{*} .
\end{equation}
Consequently, extreme operators $^{r} {\hat O}_{i}$ of
${P^{r}_{N}}^{*}$ can be combined convexly to produce all $p$-body
operators $^{p} {\hat O} \in {P^{p}_{N}}^{*}$, and hence, the
extreme points of ${P^{p}_{N}}^{*}$ can be characterized completely
by the convex combination of the extreme points of
${P^{r}_{N}}^{*}$.  More generally, convex combinations of extreme
$^{q} {\hat O}_{i} \in {P^{q}_{N}}^{*}$ generate all $p$-body
operators $^{p} {\hat O} \in {P^{p}_{N}}^{*}$ for $p < q$. Depending
upon the order of the creation and annihilation operators in $^{r}
{\hat O}_{i}$, the normal-ordered terms will have either positive or
negative coefficients.  Convex combinations of the $^{r} {\hat
O}_{i}$ can be chosen to cancel the coefficients of all terms of
degree greater than $p$.  Extreme elements are generated from the
minimum number of convex combinations to effect the cancelation.
This characterization of the set ${P^{p}_{N}}^{*}$ provides a {\em
constructive solution} of the $N$-representability problem for the
$p$-RDM.

The constructive solution---convex combinations of the operators in
Eq.~(\ref{eq:rpos})---generates the existing $N$-representability
conditions as well as new conditions.  The {\em (1,1)-positivity
conditions}~\cite{C63} are derivable from the subset of $^{r} {\hat
C}_{i}$ operators in Eq.~(\ref{eq:rpos}) of degree~1
\begin{eqnarray}
{\hat C}_{D} & = & \sum_{j}{ b_{j} {\hat a}^{\dagger}_{j} } \\
{\hat C}_{Q} & = & \sum_{j}{ b_{j} {\hat a}_{j} } .
\end{eqnarray}
Keeping the trace of the corresponding one-body operators $^{1}
{\hat O}_{D}$ and $^{1} {\hat O}_{Q}$ against the 1-RDM nonnegative
for all values of $b_{j}$ yields the conditions, $^{1} D \succeq 0$
and $^{1} Q \succeq 0$, where ${}^{1} D$ and $^{1} Q$ are matrix
representations of the 1-particle and the 1-hole RDMs and the symbol
$\succeq$ indicates that the matrix is constrained to be positive
semidefinite.

Similarly, the {\em (2,2)-positivity conditions}~\cite{GP64} follow
from considering the $^{r} {\hat C}_{i}$ operators of degree 2 in
Eq.~(\ref{eq:rpos})
\begin{eqnarray}
{\hat C}_{D} & = & \sum_{jk}{ b_{jk} {\hat a}^{\dagger}_{j} {\hat a}^{\dagger}_{k} } \\
{\hat C}_{Q} & = & \sum_{jk}{ b_{jk} {\hat a}_{j} {\hat a}_{k} } \\
{\hat C}_{G} & = & \sum_{jk}{ b_{jk} {\hat a}^{\dagger}_{j} {\hat
a}_{k} } .
\end{eqnarray}
Restricting the trace of the corresponding two-body operators $^{2}
{\hat O}_{D}$, $^{2} {\hat O}_{Q}$, and $^{2} {\hat O}_{G}$ against
the 2-RDM to be nonnegative for all values of $b_{jk}$ defines the
conditions, $^{2} D \succeq 0$,  $^{2} Q \succeq 0$, and  $^{2} G
\succeq 0$, which constrain the probabilities for finding two
particles, two holes, and a particle-hole pair to be nonnegative,
respectively.


\begin{table*}[t!]

\caption{A class of (2,4)-positivity conditions can be derived from
convex combinations of the (4,4)-positivity conditions that cancel
the 3- and 4-particle operators.  We achieve the cancelation through
tensor decomposition in the model Hamiltonians.}

\label{t:24}

\begin{ruledtabular}
\begin{tabular}{c}

(2,4)-Positivity Conditions \\

\hline

${\rm Tr}( (3 {\hat C}_{\rm xxxx} {\hat C}_{\rm xxxx}^{\dagger} +
{\hat C}_{\rm xxxo} {\hat C}_{\rm xxxo}^{\dagger} + {\hat C}_{\rm
xxox} {\hat C}_{\rm xxox}^{\dagger} + {\hat C}_{\rm xoxx} {\hat
C}_{\rm xoxx}^{\dagger} + {\hat C}_{\rm oxxx} {\hat C}_{\rm
oxxx}^{\dagger} + {\hat C}_{\rm oooo} {\hat
C}_{\rm oooo}^{\dagger}) \, {}^{2} D) \ge 0$ \\

${\rm Tr}( (3 {\hat C}_{\rm xxxo} {\hat C}_{\rm xxxo}^{\dagger} +
{\hat C}_{\rm xxxx} {\hat C}_{\rm xxxx}^{\dagger} + {\hat C}_{\rm
xxoo} {\hat C}_{\rm xxoo}^{\dagger} + {\hat C}_{\rm xoxo} {\hat
C}_{\rm xoxo}^{\dagger} + {\hat C}_{\rm oxxo} {\hat C}_{\rm
oxxo}^{\dagger} + {\hat C}_{\rm ooox} {\hat
C}_{\rm ooox}^{\dagger}) \, {}^{2} D) \ge 0$ \\

${\rm Tr}( (3 {\hat C}_{\rm xxox} {\hat C}_{\rm xxox}^{\dagger} +
{\hat C}_{\rm xxoo} {\hat C}_{\rm xxoo}^{\dagger} + {\hat C}_{\rm
xxxx} {\hat C}_{\rm xxxx}^{\dagger} + {\hat C}_{\rm xoox} {\hat
C}_{\rm xoox}^{\dagger} + {\hat C}_{\rm oxox} {\hat C}_{\rm
oxox}^{\dagger} + {\hat C}_{\rm ooxo} {\hat
C}_{\rm ooxo}^{\dagger}) \, {}^{2} D) \ge 0$ \\

${\rm Tr}( (3 {\hat C}_{\rm xoxx} {\hat C}_{\rm xoxx}^{\dagger} +
{\hat C}_{\rm xoxo} {\hat C}_{\rm xoxo}^{\dagger} + {\hat C}_{\rm
xoox} {\hat C}_{\rm xoox}^{\dagger} + {\hat C}_{\rm xxxx} {\hat
C}_{\rm xxxx}^{\dagger} + {\hat C}_{\rm ooxx} {\hat C}_{\rm
ooxx}^{\dagger} + {\hat C}_{\rm oxoo} {\hat
C}_{\rm oxoo}^{\dagger}) \, {}^{2} D) \ge 0$ \\

${\rm Tr}( (3 {\hat C}_{\rm oxxx} {\hat C}_{\rm oxxx}^{\dagger} +
{\hat C}_{\rm oxxo} {\hat C}_{\rm oxxo}^{\dagger} + {\hat C}_{\rm
oxox} {\hat C}_{\rm oxox}^{\dagger} + {\hat C}_{\rm ooxx} {\hat
C}_{\rm ooxx}^{\dagger} + {\hat C}_{\rm xxxx} {\hat C}_{\rm
xxxx}^{\dagger} + {\hat C}_{\rm xooo} {\hat
C}_{\rm xooo}^{\dagger}) \, {}^{2} D) \ge 0$ \\

${\rm Tr}( (3 {\hat C}_{\rm xxoo} {\hat C}_{\rm xxoo}^{\dagger} +
{\hat C}_{\rm xxox} {\hat C}_{\rm xxox}^{\dagger} + {\hat C}_{\rm
xxxo} {\hat C}_{\rm xxxo}^{\dagger} + {\hat C}_{\rm xooo} {\hat
C}_{\rm xooo}^{\dagger} + {\hat C}_{\rm oxoo} {\hat C}_{\rm
oxoo}^{\dagger} + {\hat C}_{\rm ooxx} {\hat
C}_{\rm ooxx}^{\dagger}) \, {}^{2} D) \ge 0$ \\

${\rm Tr}( (3 {\hat C}_{\rm xoox} {\hat C}_{\rm xoox}^{\dagger} +
{\hat C}_{\rm xooo} {\hat C}_{\rm xooo}^{\dagger} + {\hat C}_{\rm
xoxx} {\hat C}_{\rm xoxx}^{\dagger} + {\hat C}_{\rm xxox} {\hat
C}_{\rm xxox}^{\dagger} + {\hat C}_{\rm ooox} {\hat C}_{\rm
ooox}^{\dagger} + {\hat C}_{\rm oxxo} {\hat
C}_{\rm oxxo}^{\dagger}) \, {}^{2} D) \ge 0$ \\

${\rm Tr}( (3 {\hat C}_{\rm xoxo} {\hat C}_{\rm xoxo}^{\dagger} +
{\hat C}_{\rm xoxx} {\hat C}_{\rm xoxx}^{\dagger} + {\hat C}_{\rm
xooo} {\hat C}_{\rm xooo}^{\dagger} + {\hat C}_{\rm xxxo} {\hat
C}_{\rm xxxo}^{\dagger} + {\hat C}_{\rm ooxo} {\hat C}_{\rm
ooxo}^{\dagger} + {\hat C}_{\rm oxox} {\hat
C}_{\rm oxox}^{\dagger}) \, {}^{2} D) \ge 0$ \\

\end{tabular}
\end{ruledtabular}

\end{table*}

In general, the $(q,q)$-positivity conditions~\cite{EJ00,M04} follow
from restricting all $q$-body operators ${}^{q} {\hat O}$ in
Eq.~(\ref{eq:rpos}) to be nonnegative in their trace against the
$q$-RDM~\cite{M04}.  While the $(q,q)$-positive operators are not
two-body operators for $q>2$, convex combinations of them generate
two-body operators $^{2} {\hat O} \in {P^{2}_{N}}^{*}$ that enforce
the $N$-representability of the 2-RDM.  We refer to necessary
$N$-representability conditions arising from convex combinations of
$(q,q)$-positivity conditions as $(2,q)$-positivity conditions.

The simplest such constraints, the {\em (2,3)-positivity
conditions}, arise from keeping convex combinations of 3-body
operators in Eq.~(\ref{eq:rpos}) nonnegative; for example,
\begin{eqnarray}
^{2} {\hat O}_{T1} & = & \frac{1}{2} ( {\hat C}_{T1,1} \, {\hat
C}_{T1,1}^{\dagger} +  {\hat C}_{T1,2}  {\hat
C}_{T1,2}^{\dagger} ) \\
^{2} {\hat O}_{T2} & = & \frac{1}{2} ( {\hat C}_{T2,1} \,  {\hat
C}_{T2,1}^{\dagger} +  {\hat C}_{T2,2}  {\hat C}_{T2,2}^{\dagger} )
\end{eqnarray}
where
\begin{eqnarray}
 {\hat C}_{T1,1} & = & \sum_{jkl}{ b_{jkl} {\hat a}^{\dagger}_{j}
{\hat a}^{\dagger}_{k} {\hat a}^{\dagger}_{l} } \\
 {\hat C}_{T1,2} & = & \sum_{jkl}{ b^{*}_{jkl} {\hat a}_{j} {\hat a}_{k} {\hat a}_{l}} \\
 {\hat C}_{T2,1} & = & \sum_{jkl}{ b_{jkl} {\hat a}^{\dagger}_{j}
{\hat a}^{\dagger}_{k} {\hat a}_{l} } + \sum_{j}{ b_{j} {\hat a}^{\dagger}_{j} } \label{eq:T21} \\
 {\hat C}_{T2,2} & = & \sum_{jkl}{ b^{*}_{jkl} {\hat a}_{j}
{\hat a}_{k} {\hat a}^{\dagger}_{l} } + \sum_{j}{d_{j} {\hat a}_{j}}
\label{eq:T22} .
\end{eqnarray}
These conditions, known as the $T1$ and generalized $T2$ conditions
were developed by Erdahl~\cite{E78} and implemented by Zhao {\em at
al.}~\cite{P04} and Mazziotti~\cite{M04}.  In general, they
significantly improve the accuracy of the 2-positivity conditions.

Although the constructive proof given above indicates that a
complete set of $N$-representability conditions can be generated
from convex combinations of extreme elements of ${P^{r}_{N}}^{*}$,
additional conditions have not been discovered beyond the (2,2)- and
(2,3)-positivity conditions. For example, what about
(2,4)-positivity conditions---that is, $N$-representability
constraints on the 2-RDM arising from convex combinations of 4-body
operators in Eq.~(\ref{eq:rpos})?  First, we derive a class of
(3,4)-positivity conditions on the 3-RDM.

Consider the nonnegativity of the following operator ${\hat O}$
formed by the convex combination of a pair of 4-body operators from
Eq.~(\ref{eq:rpos})
\begin{equation}
\label{eq:O} {\hat O} = \frac{1}{2} ( {\hat C}_{\rm xxxx} \, {\hat
C}_{\rm xxxx}^{\dagger} + {\hat C}_{\rm xooo}  {\hat
C}_{\rm xooo}^{\dagger} )\\
\end{equation}
where the symbols ${\rm x}$ and ${\rm o}$ represent creation and
annihilation operators, respectively, in the ${\hat C}$ operators
defined as follows
\begin{eqnarray}
{\hat C}_{\rm xxxx} & = & \sum_{jklm}{ b_{jklm} {\hat
a}^{\dagger}_{j}
{\hat a}^{\dagger}_{k} {\hat a}^{\dagger}_{l} {\hat a}^{\dagger}_{m} } \\
{\hat C}_{\rm xooo} & = & \sum_{jklm}{ d_{jklm} {\hat
a}^{\dagger}_{j} {\hat a}_{k} {\hat a}_{l} {\hat a}_{m} } .
\end{eqnarray}
Importantly, the expectation value of ${\hat O}$ with $d_{jklm} =
b_{jklm}$ requires the 4-RDM because the cumulant part ${}^{4}
\Delta$ of the 4-RDM~\cite{M98,RDM} does not vanish
\begin{equation}
\sum_{jklmpqst}{ b_{jklm} b^{*}_{pqst} \, (^{4} \Delta^{jklm}_{pqst}
- ^{4} \Delta^{jqst}_{pklm} ) } \neq 0 .
\end{equation}
To obtain additional $N$-representability conditions requires that
the dependence of the ${\hat C}$ operators on the expansion
coefficients be {\em generalized from linear to nonlinear}.
Specifically, to obtain 3-RDM conditions beyond the (3,3)-positivity
constraints, we must factor the 4-particle expansion coefficients
$b_{jklm}$ and $d_{jklm}$ into products of 3- and 1-particle
coefficients $b_{j} b_{klm}$ and $b_{j} b^{*}_{klm}$ which cause the
cumulant part of the 4-RDM in $\langle \Psi | {\hat O} | \Psi
\rangle $ to vanish
\begin{equation}
\sum_{jklmpqst}{ b_{j} b_{klm} b^{*}_{p} b^{*}_{qst} \, (^{4}
\Delta^{jklm}_{pqst} - ^{4} \Delta^{jklm}_{pqst} ) } = 0 .
\end{equation}
The (3,4)-positivity condition, represented by Eq.~(\ref{eq:O}) and
the tensor decomposition of the expansion coefficients, is part of a
class of (3,4)-conditions that arises from all distinct combinations
of two 4-particle metric matrices that differ from each other in the
replacement of {\em three} second-quantized operators by their
adjoints.

A class of {\em (2,4)-positivity conditions}, shown in
Table~\ref{t:24}, can be derived from convex combinations of the
above (3,4)-positivity conditions that cancel the 3-particle
operators, that is the products of six second-quantized operators.
To effect the cancelation, the nonlinearity of the expansion
coefficients of ${\hat C}$ must be increased from $b_{j} b_{klm}$ to
$b_{j} c_{k} d_{l} e_{m}$. Specifically, the ${\hat C}$ operators in
Table~\ref{t:24} are defined as
\begin{equation}
{\hat C}_{\rm uvwz} = \sum_{jklm}{ b^{\rm u}_{j} c^{\rm v}_{k}
d^{\rm w}_{l} e^{\rm z}_{m} {\hat a}^{\rm u}_{j} {\hat a}^{\rm
v}_{k} {\hat a}^{\rm w}_{l} {\hat a}^{\rm z}_{m} } ,
\end{equation}
where ${\hat a}^{u}_{j}$ and $b^{\rm u}_{j}$ are ${\hat
a}^{\dagger}_{j}$ and $b^{*}_{j}$ if ${\rm u}={\rm x}$ and ${\hat
a}_{j}$ and $b_{j}$ if ${\rm u}={\rm o}$.  Each of the eight
(2,4)-positivity conditions in Table~\ref{t:24} generates an
additional condition by switching all ${\rm x}$'s and ${\rm o}$'s in
accordance with {\em particle-hole duality}, the symmetry between
particles and holes.  The (2,4)-conditions become the diagonal
$N$-representability conditions~\cite{E78,Cuts, D02,KM08} when $b$,
$c$, $d$, and $e$ are restricted to be unit vectors; they are more
general than the unitarily invariant diagonal conditions because
these four vectors are not required to be orthogonal.  These
(2,4)-positivity conditions are only representative of the process
by which complete conditions can be constructed from the solution of
the $N$-representability problem presented in this Letter.
Additional (2,4)-conditions in this class can be generated from
reordering creation and annihilation operators in the conditions of
Table~\ref{t:24}, and other extreme (2,4)-conditions can be
constructed from lifting the (2,3)-conditions.  A comprehensive list
of (2,4)-positivity conditions as well as (2,3)-, (2,5)-, and
(2,6)-positivity conditions, which are consistent with the
constructive solution, will be presented elsewhere~\cite{M12}.  The
(2,5)- and (2,6)-conditions include extensions of three and eighteen
classes of known diagonal conditions, respectively.

The set ${P^{2}_{N}}^{*}$ of $N$-representability conditions on the
2-RDM contains the set ${C^{2}_{N}}^{*}$ of {\em classical
$N$-representability conditions}~\cite{E78,Cuts,D02,KM08}, which
ensure that the two-electron reduced density function (2-RDF), the
diagonal (classical) part of the 2-RDM, can be represented by the
integration of a $N$-particle density function.  In different fields
the set $C^{2}_{N}$ of $N$-representable 2-RDF has been given
different names: cut polytope ~\cite{Cuts} in combinatorial
optimization and the correlation (or Boole) polytope~\cite{Cuts,P89}
in the study of 0-1 programming or Bell's inequalities.  The set
$C^{2}_{N}$, previously characterized, has important applications in
global optimization including the search for the global energy
minima of molecular clusters~\cite{KM08}, the study of classical
fluids~\cite{Fluid}, the max-cut problem in circuit design and spin
glasses~\cite{Cuts}, lattice holes in the geometry of numbers, pair
density (2-RDF) functional theory~\cite{D02}, and the investigation
of generalized Bell's inequalities~\cite{P89}.  The characterization
of the set $P^{2}_{N}$ of $N$-representable 2-RDMs represents a
significant generalization of the solution of the classical
$N$-representability problem (the Boole 0-1 programming problem). In
addition to its potentially significant applications to the study of
correlation in many-fermion quantum systems, knowledge of the set
$P^{2}_{N}$ may have important applications to ``quantum'' analogues
of problems in circuit design and the geometry of numbers.

The complete set of $N$-representability conditions firmly
solidifies 2-RDM theory as a fundamental theory of many-body quantum
mechanics with two-particle interactions.  Rigorous lower bounds to
the ground-state energy of strongly correlated quantum systems can
be computed and improved in polynomial time from subsets of the
complete $N$-representability conditions~\cite{M11} (Minimizing the
energy with a fully $N$-representable 2-RDM is a non-deterministic
polynomial (NP) complete problem because $C^{2}_{N} \subset
P^{2}_{N}$ with optimization in $C^{2}_{N}$ known to be
NP-complete~\cite{Cuts}). The present result raises challenges and
opportunities for future research that include (i) implementing the
higher $N$-representability conditions which are not in the form of
traditional semidefinite programming~\cite{M04,P04,M11}, and (ii)
determining which of the new conditions are most appropriate for
different problems in many-particle chemistry and physics.  Beyond
their potential computational applications, the complete
$N$-representability conditions for fermionic density matrices
provide new fundamental insight into many-electron quantum mechanics
including the identification and measurement of correlation and
entanglement.

\begin{acknowledgments}

The author thanks D. Herschbach, H. Rabitz, and A. Mazziotti for
encouragement, and the NSF, ARO, Microsoft Corporation, Dreyfus
Foundation, and David-Lucile Packard Foundation for support.

\end{acknowledgments}

\end{document}